\documentclass[conference]{IEEEtran}
\IEEEoverridecommandlockouts

\usepackage{amssymb}
\usepackage{graphicx}
\usepackage{tabularx}
\usepackage{tabulary}
\usepackage{url}
\usepackage{amsmath}
\usepackage[left=2cm,right=1.5cm,top=1.5cm,bottom=1.5cm]{geometry}
\usepackage{amsfonts}
\usepackage{textcomp,gensymb}
\usepackage{float}
\usepackage{threeparttable}
\usepackage{xcolor}
\usepackage{cite}
\usepackage{nomencl}
\usepackage[ruled,vlined,linesnumbered]{algorithm2e}
\usepackage{algpseudocode}
\usepackage{caption}
\usepackage{soul}

\begin{document}

\title{Deceiving Flexibility: A Stealthy False Data Injection Model in Vehicle-to-Grid Coordination}

\author{\IEEEauthorblockN{Kaan T. Gun; Xiaozhe Wang }
\IEEEauthorblockA{\textit{Department of Electrical and Computer Engineering} \\
\textit{McGill University}
}
\and
\IEEEauthorblockN{Danial Jafarigiv}
\IEEEauthorblockA{\textit{L'Institut de recherche d'Hydro-Québec (IREQ)}\thanks{This work was supported in part by the Fonds de recherche du Québec-secteur Nature et technologies (FRQ-Secteur NT) under Grant 2023-NOVA-314338 and in part by the Hydro-Québec, the Institut de Valorisation des Données (IVADO), MITACS under Grant IT27493.}
}
}

% \author{\IEEEauthorblockN{Kaan T. Gun; Xiaozhe Wang }
% \IEEEauthorblockA{\textit{Department of Electrical and Computer Engineering} \\
% \textit{McGill University}
% Montreal, Canada \\
% \{kaan.gun, xiaozhe.wang2\}@mcgill.ca}
% \and
% \IEEEauthorblockN{Danial Jafarigiv; Marthe Kassouf}
% \IEEEauthorblockA{\textit{L'Institut de recherche d'Hydro-Québec (IREQ)} \\
% Montreal, Canada \\
% \{jafarigiv.danial2, kassouf.marthe\}@hydroquebec.com}
% }

\maketitle

\vspace*{-3em}

% Write abstract here
\begin{abstract}
Electric vehicles (EVs) in Vehicle-to-Grid (V2G) systems act as distributed energy resources that support grid stability. Centralized coordination such as the extended State Space Model (eSSM) enhances scalability and estimation efficiency but may introduce new cyber-attack surfaces. This paper presents a stealthy False Data Injection Attack (FDIA) targeting eSSM-based V2G coordination. %\color{red}
Unlike prior studies that assume attackers can disrupt physical charging or discharging processes, we consider an adversary who compromises only a subset of EVs, and limiting their influence to the manipulation of reported State of Charge (SoC) and power measurements.\color{black}
% Unlike prior studies that assume attackers can disrupt physical charging or discharging processes, we consider a limited adversary capable of compromising only a subset of EVs and manipulating their reported State of Charge (SoC) and power measurements.
By doing so, the attacker can deceive the operator’s perception of fleet flexibility while remaining consistent with model-based expectations, thus evading anomaly detection. Numerical simulations show that the proposed stealthy FDIA can deteriorate grid frequency stability even without direct access to control infrastructure. These findings highlight the need for enhanced detection and mitigation mechanisms tailored to aggregated V2G frameworks.
%This paper addresses a critical cybersecurity vulnerability
%in Vehicle-to-Grid (V2G) systems: stealthy attacks that lead to an
%overestimation of grid flexibility and compromise frequency stability.
%Unlike existing research focusing on overt attacks causing physical
%damage, we propose a novel Man-in-the-Middle (MitM) False Data Injection
%Attack (FDIA) model tailored for V2G environments where operators
%utilize an aggregated EV model, specifically the extended state-space
%model (eSSM), for decision-making. Our approach involves an attacker
%compromising a subset of EVs to manipulate cts. To maintain stealth, the attack design
%adheres to eSSM conditions by limiting distributional deviations and
%ensuring consistency in individual EV data. We formulate a two-stage
%optimization problem to design group attack vectors (manipulated SoC,
%power, and percentage of EVs) and assign them to individual vehicles,
%maximizing the discrepancy between true and estimated flexibility
%while remaining undetected. Numerical experiments validated the
%attack's impact on fleet operator's estimated flexibility and its
%subsequent integration into a 2-area Automatic Generation Control
%(AGC) system to demonstrate the deterioration of grid frequency
%stability when EVs fail to provide adequate power support due to
%misestimated flexibility. These findings underscore the urgent need
%for robust defense mechanisms in future V2G systems.

\end{abstract}

% \makenomenclature

% \nomenclature{$\boldsymbol{E}$}{Transformation matrix}
% \nomenclature{$t_{s,i} , t_{f,i}$}{Charge start and finish times}
% \nomenclature{$S_{s,i}, S_{d,i}$}{State and departure SoC for EV $i$}
% \nomenclature{$\eta_{s,i}, \eta_{d,i}$}{Charge and discharge efficiency $i$}
% \nomenclature{$\boldsymbol{x}(k)$}{Aggregated state distribution, i.e., the proportion of EVs located in the corresponding SOC intervals across operation modes \color{black}}
% \nomenclature{$Q_i$}{Battery capacity of EV $i$}
% \nomenclature{$S_{min}, S_{max}$}{SoC range}
% \nomenclature{$k$}{Simulation time index}
% \nomenclature{$i$}{Index of EVs}
% \nomenclature{$N_p$}{Number of $T$ intervals in $T_p$}
% \nomenclature{$N_s$}{Number of SoC intervals}
% \nomenclature{$\boldsymbol{\pi}_i$}{%\color{blue}
% EV's probability to change state indexes} %\color{red} do you mean the EV's probability to change state indexes? I think the current description is not clear enough \color{black}}
% \nomenclature{$P_{c,i} , P_{d,i}$}{Rated charging and discharging power}
% \nomenclature{$S_{s,i}, S_{d,i}$}{Start and departure SoC}

% \printnomenclature

\section{Introduction}
%\color{red}
%1. EVs are increasing, there is an opportunity of V2G to enhance and support the grid (a few examples).

Electric vehicles (EVs) are gaining attention as governments worldwide implement supportive policies and invest in green technologies to achieve emission reduction goals \cite{iea2024evoutlook}. One promising advancement in this domain is Vehicle-to-Grid (V2G) technology, which enables bidirectional energy flow between EVs and the power grid, allowing EVs to act as distributed energy resources (DERs) that enhance grid stability \cite{6123186} and facilitate the integration of renewable energy \cite{V2GReview, MultiObjectiveOptimalScheduling, V2GFacilityforSmart_Grid, YANG20242786}.

%\color{red}
%2. Introduce previous works (centralized algorithms) how in v2G framework, the EVs can support the grid. mention the challenge is that communication cost or the model size will increase as the number of EVs increase. To address this, SSM model was proposed (one sentence to explain why). Because of these advantages, SSM holds the potential to be
%implemented widely in practice. May be one sentence to describe the benefits.

%\color{green}

However, challenges persist due to the uncertainty of travel patterns, diverse charging behaviors, and the complexity of managing large EV fleets. In V2G frameworks, centralized algorithms have been proposed where a central operator coordinates EV charging to support the grid and optimize fleet performance \cite{8727484, hierarchicalOptADMM}. A key challenge with such approaches is scalability; as the number of EVs grows, the communication cost and computational burden associated with collecting frequent state updates (like State-of-Charge (SoC)) from each vehicle increase significantly \cite{ChifuRL}. To address this limitation, the State Space Model (SSM) was introduced \cite{8764460} and further extended in \cite{Liu_Wang_2024}. Given that V2G systems rely on sophisticated communication and control, research into the cybersecurity vulnerabilities of charging infrastructure has intensified in \cite{SANDIA} to ensure the reliability and safety of grid operations.
%\color{red}

%3. There are pervious works considering cyber atttacks targeting EVs.  %under the V2G framework.
%Review all the previous cybersecurity works on EVs. Mention that if brutal/ cruel attacked laughed in the V2G framework, it will be easily detected because operator has their model running in the background. Anomalies can be easily detected if the received measurements are different from the expected values.

%the gap. All these works don't consider V2G and therefore  there is no assumed EV models for the operator. assume attackers have direct control over the vehicles or chargers, or assume the attacker knows the network and dynamic model parameters of the grid, which are not practical.

%\color{blue} \cite{SANDIA}.

% \color{red} add a connecting sentence explaining why cybersecurity is important in V2G. \color{black}

 % \color{black} Previous cybersecurity studies have explored threat assessments on spoofing, tampering, and Distributed Denial-of-Service (DDoS) in EV chargers \cite{10020644,Detailed_Security_Assessment} and, in smart charging management systems \cite{TASNIM2025,Remote_Code}. While others have investigated designing attack vectors targeting individual EV charging sessions, potentially causing physical damage to vehicle batteries \cite{8960519}.

 %\color{red}
 Previous cybersecurity studies have explored threat assessments on spoofing, tampering, and Distributed Denial-of-Service (DDoS) in EV chargers \cite{10020644,Detailed_Security_Assessment}; similarly, in the context of smart charging management systems\cite{TASNIM2025,Remote_Code} others have investigated designing attack vectors targeting individual EV charging sessions that could potentially cause physical damage to vehicle batteries \cite{8960519}. \color{black}Furthermore, the integration of EVs as cyber-physical assets has prompted investigations into grid stability under large-scale load manipulation attacks \cite{9091609}, often assuming the attacker gains significant or full control over charging assets.

However, when considering attacks within an actively managed V2G framework, the presence of a central coordinator changes the landscape. Such coordinators typically employ operational models to predict and manage the fleet's aggregated behavior for grid services. Consequently, overt or large-magnitude brute-force attacks, which would cause significant discrepancies between the measurements and the values expected by the operator's model, are likely to be rapidly identified through anomaly detection mechanisms. In contrast to prior works \cite{SANDIA,Remote_Code,8960519,9091609}, which often assume direct attacker control over individual chargers or EVs, or do not explicitly account for the monitoring capabilities inherent in centralized V2G coordination, we consider a more nuanced scenario. Our model assumes the fleet operator employs the extended State Space Model (eSSM) as a centralized V2G control strategy \cite{8764460,Liu_Wang_2024}, while the attacker has limited knowledge and lacks direct control over the charging infrastructure, necessitating more sophisticated or stealthy approaches to impact the system effectively.

%\color{red}
%4. In this paper, we investigate the cyberattack model in V2G framework. We assume that the operator has a model running to monitor and control the charging dynamics of EVs. Particularly, given the potential of SSM, we assume the model is SSM. In this framework, we also assume the attackers have limited knowledge of the system and cannot directly alter the control signals sent to individual EVs. We propose an FDIA attack model that can stealthily do what......deteriorate the grid frequency.......
%. We hope to raise the attention from community to further study cybersecurity....
%show that

%cannot to avoid being detected, the attacker cannot detect
%the SSM model is being deployed for V2G because of its scalability, monitoring and estimating capability, enhanced situational awareness? Although SSM is more robust to cyberattacks because of the reduced communication requirements....there are still possibilities to construct cyberattacks based on the SSM model. We will show 1) attackers don't have to have the control over EVs or chargers to launch a stealthy cyberattack that can deteriorate the grid's frequency 2)

Such strategy allows for monitoring and controlling the charging dynamics of the fleet, leveraging the eSSM's recognized advantages in scalability, communication efficiency, and state estimation, which enhances the situational awareness\cite{8764460,Liu_Wang_2024}. While eSSM-based control inherently offers some resilience against certain threats due to reduced communication needs and aggregated modeling, possibilities remain for constructing sophisticated cyberattacks. In this work, we specifically analyze a scenario under the realistic assumption that attacker cannot directly compromise the V2G framework or alter the control signals dispatched to individual EVs or chargers. We propose a stealthy Man-in-the-Middle (MitM) False Data Injection Attack (FDIA) model targeting the eSSM-based V2G system. This attack involves compromising a subset of participating EV  manufacturers' backend communication and manipulating their reported data with the aim of misleading the operator's state estimation and control logic. We demonstrate that even without accessing the charger or EV infrastructure, an attacker can exploit local information to coordinate the injection of false data stealthily, thereby avoiding straightforward detection by the central coordinator's anomaly detection mechanisms. The potential consequence of such a successful stealthy attack is the degradation of grid services, possibly leading to operational inefficiencies, financial losses, or impacting grid frequency.

\section{Preliminaries}\label{Preliminaries}

This section introduces the individual and aggregated EV model, detailing their SoC dynamics. A key advantage of the eSSM is that it scales efficiently with the number of vehicles, as its computational complexity is independent of fleet size and instead depends solely on the granularity of the discrete SoC levels. It also addresses communication bottlenecks by broadcasting a unified control signal while requiring EVs to send less frequent but periodic measurements.
Thus, the operator utilizes its model to infer the aggregated state and power flexibility of fleets, enabling the utility to dispatch power within the bounds of the fleets’ capabilities. Its high accuracy, computational efficiency, and reduced communication overhead make the eSSM highly suitable for managing large-scale EV fleets in grid services.

% \subsection{State Model of an Individual Electric Vehicle}
\vspace*{-3pt}
\subsection{Individual Electric Vehicle Model (IEVM)}\label{IEVM}

An individual EV ($i$) operates within a well-defined set of parameters (charge start and finish times ($t_{s,i}, t_{f,i}$), start and departure SoC ($S_{s,i}, S_{d,i}$), and SoC range ($S_{min}, S_{max}) $) when connected to a charging station. The fleet operator
% \color{green} This is explained in Fig. \ref{fig:Communication Flow} \color{red} We may need to make operator, power utility, and users clearly \color{black}
maintains additional key characteristics such as battery capacity \( Q_i \), and power limits \( P_i \in \{P_{c,i}, P_{d,i}\}\), along with the efficiency ($\eta_{i}$).

There are three operating modes for an EV: Charging Mode (CM), where it draws power from the grid to increase its SoC; Idle Mode (IM), where it remains connected but does not exchange power; or Discharging Mode (DM), where it delivers power back to the grid. For the EV index ($i$), the SoC dynamics can be expressed in discrete time, where its SoC transitions depend on the power exchange with the grid. Mathematically, the SoC at time step \( k+1 \) is determined as;

\vspace*{-8pt}
\begin{equation}
    \label{eq1}
    S_i(k+1) =
        \begin{cases}
        S_i(k) - P_i(k) \cdot \eta_{i} \cdot \frac{T}{Q_i}, & P_i(k) < 0, \text{CM} \\
        S_i(k), & P_i(k) = 0, \text{IM} \\
        S_i(k) - \frac{P_i(k)}{\eta_{i}} \cdot \frac{T}{Q_i}, & P_i(k) > 0, \text{DM}
        \end{cases}
\end{equation}
\noindent The simulation time index is expressed as $k$. This ensures the feasible operational region of an EV is constrained by its SoC and power limits, and the battery remains within its predefined charge boundaries.
% \color{red} Is Fig. 1 still needed? \color{black}
% \begin{figure}[h]
%     \centering
%     \vspace*{-8pt}
%     \includegraphics[width=0.40\textwidth]{figs/operationarea.png}
%     \caption{Individual Vehicle Operation Area}
%     \label{fig:OperatingArea}
% \end{figure}

%\st{Connected vehicles operational area is represented by the shaded region in Fig. \ref{fig:OperatingArea}. The upper boundary A-B represents the highest possible SoC trajectory during charging, while the lower boundary A-D corresponds to the lowest SoC trajectory due to discharging. The segments B-C and D-E mark the upper and lower SoC limits when the vehicle is idle, meaning no energy is being transferred. Additionally, a forced charging period ensures the vehicle reaches its required SoC before departure. This operational area guarantees that each EV can participate in grid flexibility services while adhering its SoC and power limits and reaching its required charge levels before departure.}

\vspace*{-5pt}
\subsection{The Extended State Space Model of Aggregated EVs}\label{The Extended State Space Model of Aggregated EVs}

The State Space Model (SSM) represents the fleets distribution across discrete states (i.e.  the proportion of EVs located in the corresponding SoC intervals across operation modes) $\boldsymbol{x} \in \mathbb{R}^{3N_{s}}$. For each operation mode (CM, IM, DM), the SoC range \([S_{min}, S_{max}]\) is partitioned into number of discrete SoC intervals \(N_{s}\). All connected EVs are represented at an index of the aggregated state $\boldsymbol{x}$ according to their power and SoC measurements.
% The index state of all connected EVs is determined by their measurements, and each is placed to its corresponding index in $\boldsymbol{x}$.
This formulation enables the development of a Markov Transition Model (MTM) $\boldsymbol{A} \in \mathbb{R}^{(3N_{s}, 3N_{s})}$, which describes the dynamic evolution of the aggregated EV population. Consequently, the state distribution at the next time step is expressed as $\boldsymbol{x}(k+1) = \boldsymbol{A} \boldsymbol{x}(k)$. It is assumed that EVs are uniformly distributed within each state interval \cite{9042884}.

\color{black}

Beyond the standard operational states, additional conditions affect the ability of EVs to provide grid flexibility. Specifically, when a vehicle reaches its SoC limits ($S_{max,i}, S_{min,i}$), its ability to contribute to charging or discharging services becomes asymmetric. The extended State Space Model (eSSM) captures this effect by introducing additional \emph{Special States} indexes $SS_{MAX}\!=\!{3N_s+1}$ and $SS_{MIN}\!=\!{3N_s\!+\!3}$ \cite{Liu_Wang_2024}. Furthermore, EVs that require immediate charging to meet their departure SoC target transition into a Forced Charging State index ($FCS\!=\!{3N_s+2}$), effectively removing them from grid participation.
\begin{figure}[h]
    \centering
    \vspace*{-10pt}
    \includegraphics[width=0.50\textwidth]{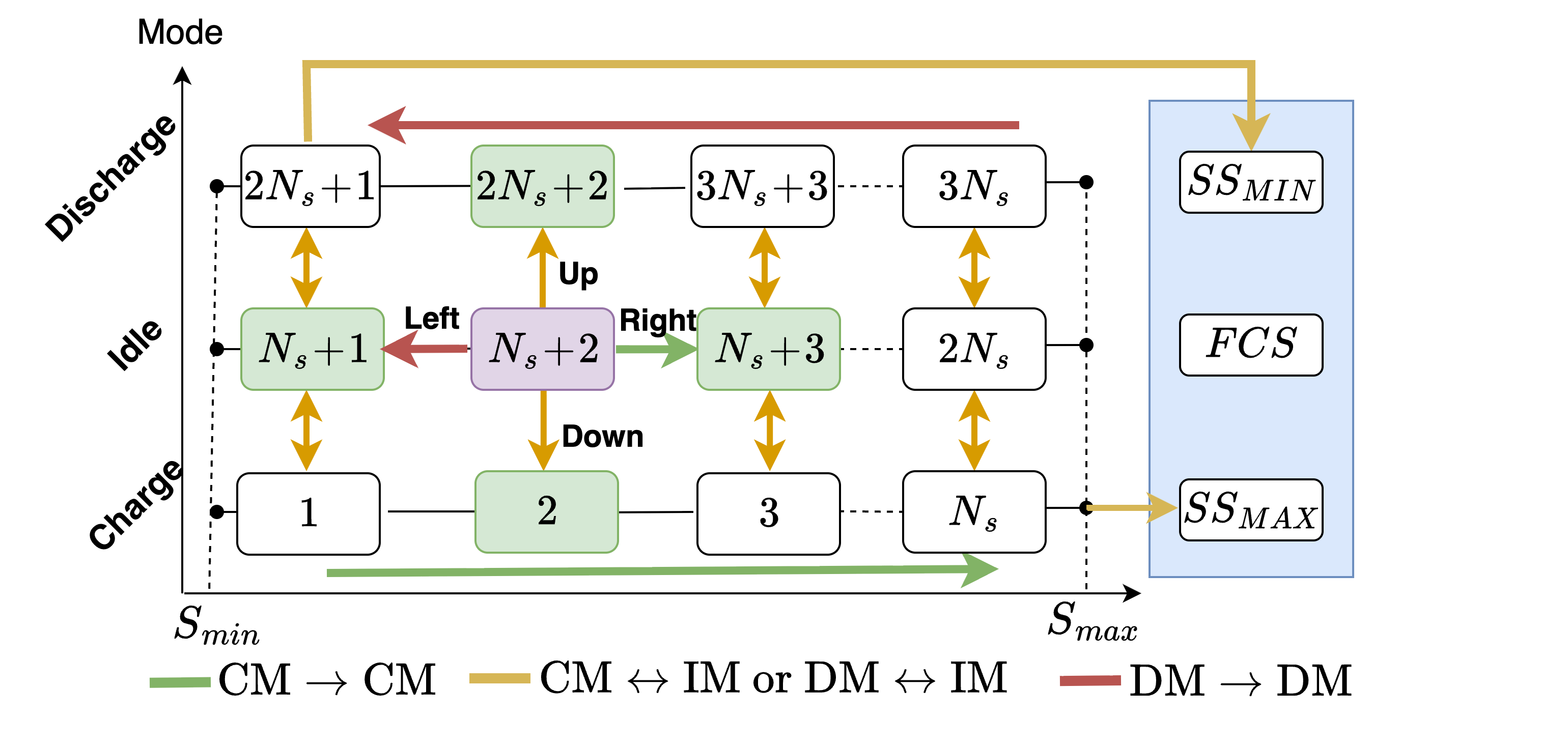}
    \caption{Illustration of State Indexes and Transitions. }
    \label{fig:StateTransitions}
    \vspace*{-10pt}
\end{figure}

By incorporating these special state indexes, \(\boldsymbol{x}\) expands to a dimension of \(3N_{s}\!+\!3\), and \(\boldsymbol{A} \in \mathbb{R}^{(3N_{s} + 3, 3N_{s} + 3)}\). %\color{red}
As represented in Fig. ~\ref{fig:StateTransitions}, \color{black}this extended formulation ensures that the eSSM accurately represents the dynamic state distribution of the fleet while accounting for constraints that impact grid flexibility.

The aggregated power output $y(k)$ is calculated based on the distribution $\boldsymbol{x}(k)$:

\vspace*{-3pt}
\begin{equation}
    \label{eq_y}
    y(k) = P_{ave}(k)  \mathcal{N}(k) \cdot \boldsymbol{D} \cdot \boldsymbol{x}(k)
\end{equation}

\noindent
where $\boldsymbol{D} = \begin{bmatrix} -\mathrm{\boldsymbol{1}}_{1\times N_s} & \mathrm{\boldsymbol{O}}_{1 \times N_s} &
\mathrm{\boldsymbol{1}}_{1 \times N_s} &
0 & -1 & 0\end{bmatrix} ^\intercal$ and $P_{ave}$ is the statistical expectation of the charging powers $P_{c}$ of the number of all connected EVs $\mathcal{N}(k)$, which can be calculated as $
P_{\mathrm{ave}}
= \int_{P_{c,\min}}^{P_{c,\max}}
  P_{c}\,f_{c}(P_{c})\,\mathrm{d}P_{c},
$
where $P_{c, min}$, $P_{c, max}$ are the bounds of is the probability density function (PDF) $f_c(P_c)$.

%\color{red}
The PDF characterizes the diversity of charging power ratings within the population. Because the fleet composition is dynamic, the PDF varies over time as the connection states of individual EVs change. By leveraging real-time data points from connected vehicles, the operator recalculates $f_c(P_c)$ at each control interval $T_{p}$, thereby providing a more precise $P_{ave}$ estimate for the aggregated power calculation in \eqref{eq_y}. \color{black}

%\color{blue}

\begin{figure}[h]
    \centering
    \vspace*{-13pt}
    \includegraphics[width=0.45\textwidth]{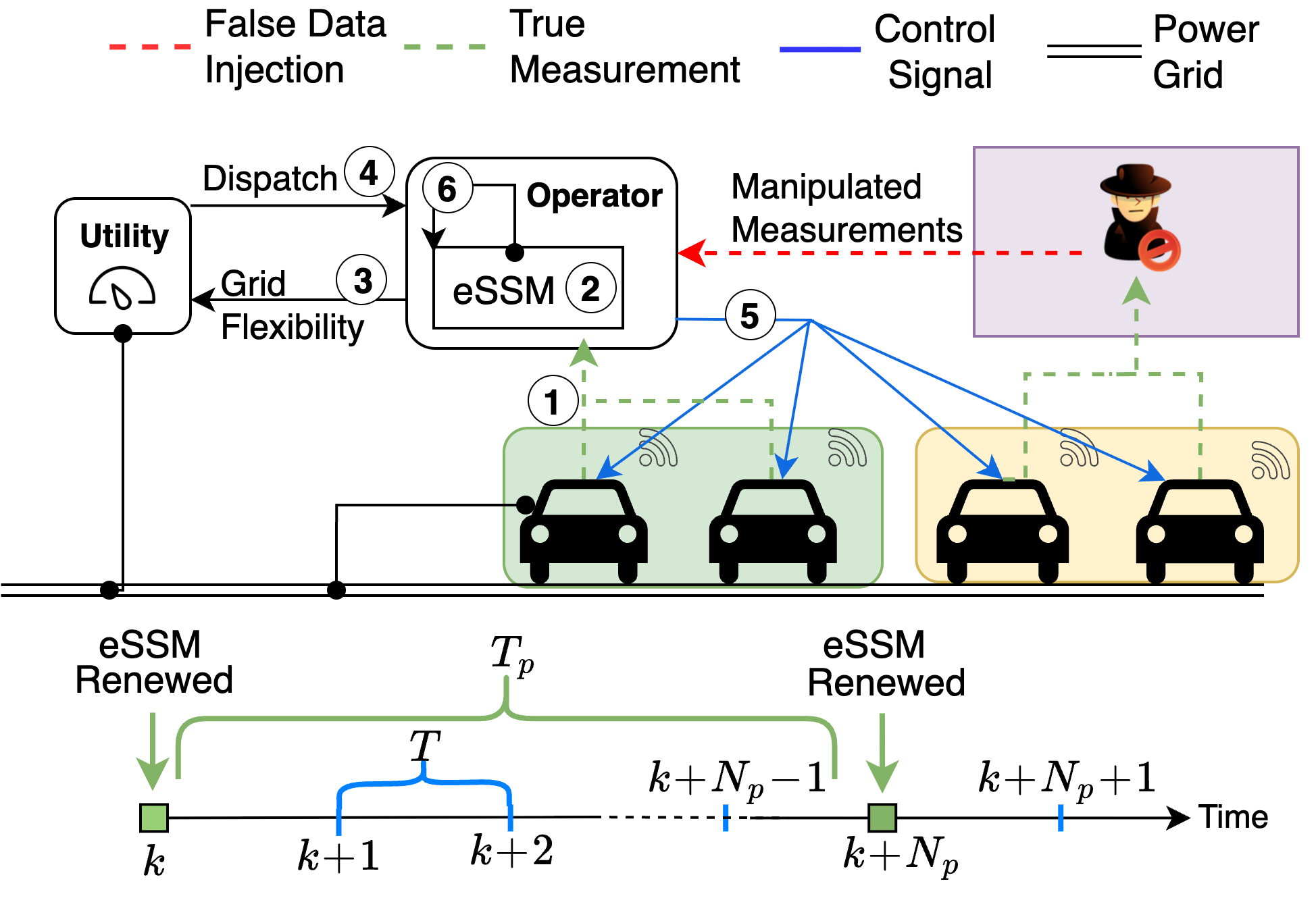}
    % \caption[Communication Flow]%
    % { Communication Flow \par \small Compromised EVs are displayed in a yellow box, while safe EVs are shown in a green box.}
    \caption{Overview of V2G Communication and eSSM execution timeline}
    \label{fig:Communication Flow}
    \vspace*{-5pt}
\end{figure}

Fig.~\ref{fig:Communication Flow} describes the operational flow of the V2G framework, which proceeds in the following steps:
\begin{enumerate}
    \item \textbf{Data Transmission:} Once an EV connects to a charger, it transmits user-defined constraints ($t_s, t_f, S_c, S_d$) and current measurements ($S(k), P(k)$) to the system operator. Subsequently, these measurements are updated every $T_p$ interval.
    \item \textbf{Model Construction:} The operator uses the aggregated data from all vehicles to construct an eSSM.
    \item \textbf{Flexibility Reporting:} The eSSM estimates the total available grid flexibility and sends its upper and lower bounds to the utility.
    \item \textbf{Power Dispatch:} In response, the utility dispatches a power setpoint to the operator.
    \item \textbf{Model Update:} The eSSM computes feedback control signals and updates the eSSM to estimate its state at the next interval ($k+1$) to maintain model accuracy.
    \item \textbf{Control Command Broadcasting:} The feedback control signals ($\boldsymbol{u}, \boldsymbol{v}$) are broad-casted to all EVs.
\end{enumerate}

It is assumed that the operator has a prior knowledge of all vehicle specifications ($Q, S_{min}, S_{max}, P_{min}, P_{max}, \eta$), which remain constant. As depicted in Fig.~\ref{fig:Communication Flow}, the eSSM updates its internal state at each time step $k$ over a horizon of $T$ intervals. The model is considered \emph{stale} after it has been updated $N_p$ times, at which point it is renewed with the most recent EV measurements. Thus, eSSM is renewed every $T_p\!=\!T N_p$ interval.

% \begin{figure}[h]
%     \centering
%     \vspace*{-10pt}
%     \includegraphics[width=0.40\textwidth]{figs/timeline.png}
%     \caption[eSSM Operation Timeline]%
%     {eSSM Operation Timeline}
%     \label{fig:timeline}
% \end{figure}
% \vspace*{-10pt}

 % 1) Vehicle measurements sent to the operator every $T_P$ interval. 2) Historic vehicle measurements used to construct the Markov Transition Matrix while recent vehicle measurements are used to construct aggregated vehicles state distribution $\boldsymbol{x}$ for the eSSM model. 3) The eSSM model sends the upper and lower bounds of the estimated grid flexibility. 4) Every $T$ interval model estimates its new state. 5) Utility requests a dispatch $P_r$. 6) SSM constructs the feedback signals and broadcasts the control signal to all vehicles.

%\st{EVs communicate with operators in a dual-layer approach: while vehicle measurements are sent periodically, operators broadcast real-time control signals. }

%\color{blue}
% \subsection{Grid Flexibility}

The upper and lower boundaries of the fleets grid flexibility is denoted as ${y_u(k)}$ and $y_l(k)$ is calculated based on $\boldsymbol{x}(k)$, respectively:
% (\ref{eq_y_ul})

\vspace*{-10pt}
\begin{equation}
\begin{aligned}
    \label{eq_y_ul}
    y_u(k) = P_{ave}(k) \mathcal{N}(k) \cdot \boldsymbol{D}_{u} \cdot \boldsymbol{x}(k) ,\\
    y_l(k) = P_{ave}(k) \mathcal{N}(k) \cdot \boldsymbol{D}_l \cdot \boldsymbol{x}(k) ,\\
\end{aligned}
\end{equation}

\noindent
where $\boldsymbol{D}_u = \begin{bmatrix} \mathrm{\boldsymbol{O}}_{1 \times N_s} &
    \mathrm{\boldsymbol{1}}_{1 \times N_s}&
    \mathrm{\boldsymbol{2}}_{1 \times N_s}&
    0 &
    0 &
    2
    \end{bmatrix} ^\intercal $,
    $
    \boldsymbol{D}_l = \begin{bmatrix}
    \mathrm{\boldsymbol{2}}_{1 \times N_s}&
    \mathrm{\boldsymbol{1}}_{1 \times N_s}&
    \mathrm{\boldsymbol{O}}_{1 \times N_s}&
    2& 0& 0
    \end{bmatrix} ^\intercal$

% \vspace*{-10pt}
% \begin{gather*}
%     \boldsymbol{D}_u = \begin{bmatrix} \mathrm{\boldsymbol{O}}_{1 \times N_s} &
%     \mathrm{\boldsymbol{1}}_{1 \times N_s}&
%     \mathrm{\boldsymbol{2}}_{1 \times N_s}&
%     0 &
%     0 &
%     2
%     \end{bmatrix} ^\intercal
%     \\
%     \boldsymbol{D}_l = \begin{bmatrix}
%     \mathrm{\boldsymbol{2}}_{1 \times N_s}&
%     \mathrm{\boldsymbol{1}}_{1 \times N_s}&
%     \mathrm{\boldsymbol{O}}_{1 \times N_s}&
%     2& 0& 0
%     \end{bmatrix} ^\intercal
% \end{gather*}

To effectively regulate the participation of EVs in grid services, a feedback control mechanism is integrated into the eSSM.
The feedback signal governs EV transitions between operating modes in response to external grid conditions. This ensures that the aggregated EV population dynamically adjusts its charging behavior while maintaining system stability.

% \subsection{Control}\label{Control}

The feedback signal, denoted as \( \boldsymbol{u}(k), \boldsymbol{v}(k) \in [-1,1]^{N_s+3}\) are introduced \eqref{eq3} to influence the state distribution at each time step.
\begin{equation}
    \label{eq3}
    \boldsymbol{x}(k+1) = \boldsymbol{A} \boldsymbol{x}(k) + \boldsymbol{B} \boldsymbol{u}(k) + \boldsymbol{C} \boldsymbol{v}(k),
\end{equation}
where
$
\boldsymbol{B} =  \begin{bmatrix}-\mathrm{\boldsymbol{I}}_{N_s \times N_s}& \mathrm{\boldsymbol{I}}_{N_s \times N_s}& \mathrm{\boldsymbol{O}}_{N_s \times N_s} \end{bmatrix}^\intercal
$,
$
\boldsymbol{C} =  \begin{bmatrix}\mathrm{\boldsymbol{O}}_{N_s \times N_s}& -\mathrm{\boldsymbol{I}}_{N_s \times N_s}& \mathrm{\boldsymbol{I}}_{N_s \times N_s}  \end{bmatrix}^\intercal
$ excluding the special states, are the constant matrices that determines how the feedback signal modifies the state transitions across different SoC intervals.

$\boldsymbol{u}(k), \boldsymbol{v}(k)$ vectors are derived based on $\boldsymbol{x}(k)$ and designed to enforce grid-supporting behaviors while respecting individual EV directives. If the total power requested by the utility $\Delta{P_r}$ is within the bounds of grid flexibility, the operator constructs the feedback signals based on a state-based priority function $\rho(x(k), \mathcal{N}, P_{ave})$. e.g. when power reduction is requested, EVs that are in CM and have high-SoC values are prioritized for switching to IM.

Finally, \( \boldsymbol{u}(k), \boldsymbol{v}(k) \) are mapped to control signals $\boldsymbol{u_s}$ and $\boldsymbol{v_s}$ that consist of two probabilistic vectors $[0,1]^{N_s}$, along with a Control Direction Element $\text{CDE}\in\; \{-1,\;1\} $, instructing EVs on whether to transition between operating modes in response to grid requirements. Each EV autonomously determines whether to execute the requested transition based on where it locates in the state space (see Fig. \ref{fig:StateTransitions}).

\vspace*{-10pt}
\begin{equation}
    \label{eq:CDE}
    \small
\boldsymbol{u_s} :
  \begin{cases}
    \text{CM} \rightarrow \text{IM}  \quad \text{if CDE} +\\
    \text{CM} \leftarrow \text{IM}   \quad \text{if CDE} -
  \end{cases}
\quad
\boldsymbol{v_s} :
  \begin{cases}
    \text{IM} \rightarrow \text{DM}   \quad \text{if CDE} +\\
    \text{IM} \leftarrow \text{DM}   \quad \text{if CDE} -
  \end{cases}
\end{equation}

These control signals represents the probability distribution of switching at each state index, and are then broadcast to all EVs. Since individual vehicles know their current state index, corresponding vehicle switch their operation modes, resulting the aggregated change in power flow $\Delta{P_{EV}}\approx \Delta{P_r}
$ \cite{8764460}. This approach eliminates the need for the operator to generate individual control signals for each vehicle.

% Main Part
\section{The proposed stealthy FDIA model} %Methodology}

We assume an attacker performs a MitM-FDIA by compromising a random subset of vehicles. As depicted in Fig. \ref{fig:Communication Flow}, the attack targets the vehicle-to-operator measurement pathway by accessing  backend servers of certain EV manufacturers. This allows the adversary to manipulate the SoC and power measurements from compromised EVs over time, after the initial transmission. The attacker has no control over the physical charging process itself, unable to  vary power levels, physically disconnect EVs, or alter the true energy stored in the battery. %\color{blue}
Its objective is to deviate operators view of the estimated grid flexibility from true flexibility.
\color{black}

% \begin{figure}
%     \centering
%     \includegraphics[width=0.45\textwidth]{figs/overall_attack_map.png}

%     \caption[Attack Sequence]%
%     { Attack Sequence \quad \small All EV measurements are sent to the operator every $T_p$ interval. Control signal and grid measurements are sent every $T$ interval.  }
%     \label{fig:Attack Sequence}
% \end{figure}

\subsection{Stealthy Conditions} \label{Stealthy Conditions}

Since the state distribution $\boldsymbol{x}$ is estimated using eSSM over a period $T_p$, both $\boldsymbol{x}$ and $A$ are updated based on the most recent EV measurements. Let \(\boldsymbol{x}_{\text{eSSM est.}}(k)\) denote the last estimated state at the end of $T_p$ by the eSSM (i.e., the stale state distribution). When the eSSM is updated with new, real-time EV measurements, its accuracy dictates that the new state vector $\boldsymbol{x}_{\text{true}}(k)$ should closely align with the prior estimate. Therefore, to ensure stealthy conditions, the eSSM's estimates derived from manipulated measurements must closely approximate those based on real-time measurements.

% When the eSSM model is updated with most up-to-date EV measurements from real life, since the eSSM is accurate, the \emph{new} state vector \(\boldsymbol{x}_{\text{new}}(k)\) should be close to the estimate. In other words, to ensure stealthy, the eSSM estimates based on the manipulated measurements should be close to the real-life true measurements.
% \color{red} update the eqn, may polish the previous writing\color{black}

\vspace*{-13pt}
\begin{equation}
    \label{eq:StealthCondition}
    \left|\boldsymbol{x}_{\text{true}}(k) - \boldsymbol{x}_{\text{eSSM est.}}(k)\right| < \epsilon, \quad \text{when } \text{mod}(k, N_p) = 0
\end{equation}

The aggregate stealth bound in \eqref{eq:StealthCondition} limits large deviations in the fleet's state vector. However, since the operator receives measurements from individual EVs, the reported values must also respect each EV's SoC and power constraints over the period $T_p$:
\vspace*{-6pt}
\begin{equation}
\label{eq:ind_ev_SoC_condition}
    P_{d,i}T_p \;\le\; Q_i \cdot  sgn(\Delta S_i )\lceil | \Delta S_i  |\rceil \;\le\; P_{c,i}T_p
\end{equation}
\begin{equation}
\label{eq:ind_ev_P_condition}
    P_{d,i} \;\le\; P_i' \;\le\; P_{c,i}
\end{equation}
\noindent
where $\Delta S_i=S_i'-S_i$ is the change in SoC for EV $i$ and determined by the successive measurements denoted as ($S'_i$, $ P_i'$). Round away from zero is used due to limited granularity of reported SoC ($G_{SoC} =\%1$).

\subsection{Multi-Stage FDI Attack} \label{Attackers Simulation}

\begin{figure}[h]
    \centering
    \vspace*{-8pt}
    \includegraphics[width=0.48\textwidth]{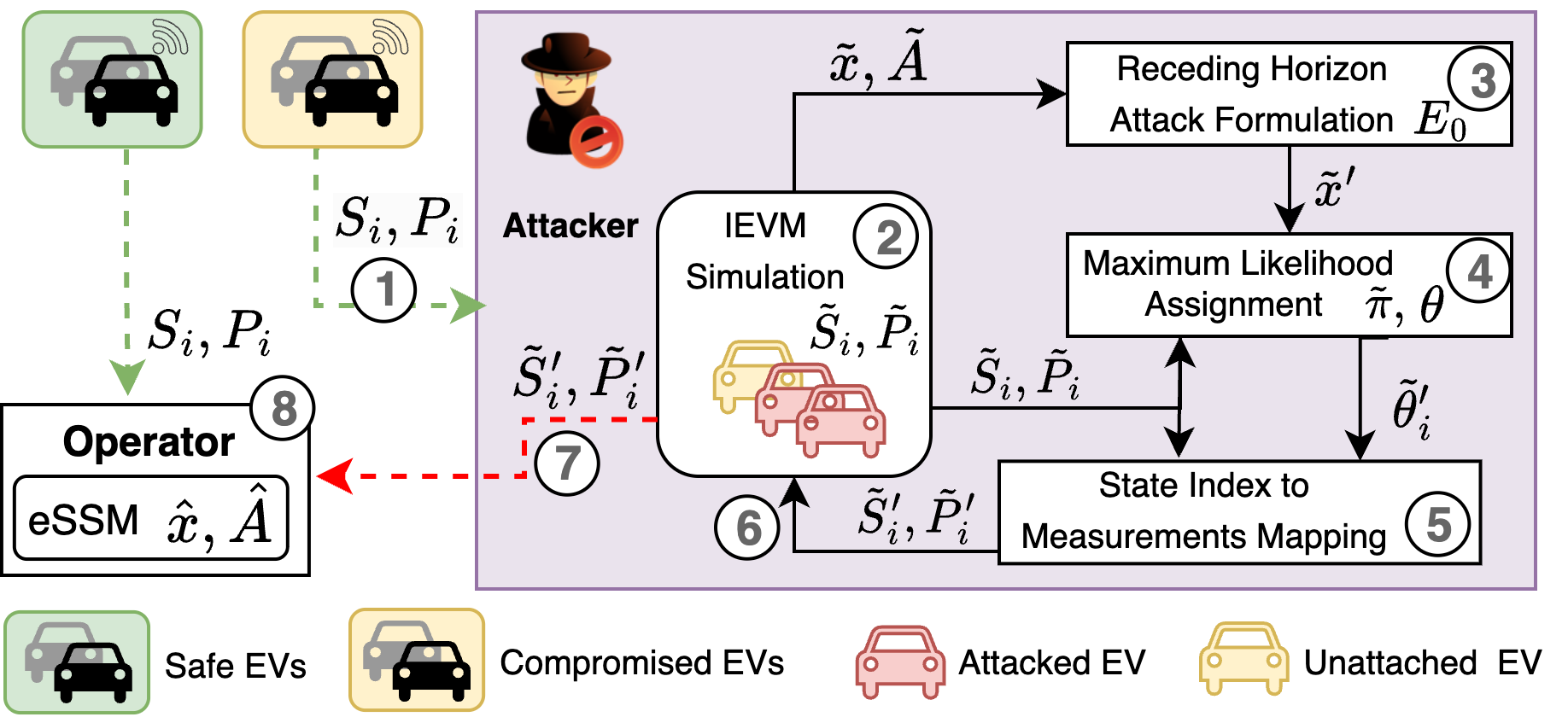}
    \vspace*{8pt}
    \caption[FDI Attack Procedure \color{red}FDI or FDIA, be consistent\color{black}]%
    {FDI Attack Procedure \quad \small Green and red dashed lines indicate the true and manipulated EV measurements respectively.}
    \label{fig:Attack_loop_inner}
    \vspace*{-5pt}
\end{figure}

%\color{blue}
The attacker optimizes its FDIA strategy by exploiting the dynamics of the eSSM. To remain undetected, the attacker ensures that both the previously transmitted individual EV measurements, and the newly manipulated measurements, satisfy the stealthy constraints defined in \eqref{eq:ind_ev_SoC_condition} and \eqref{eq:ind_ev_P_condition}. The attack optimization is executed at the same frequency, i.e., every $T_p$, as the EV measurement transmission frequency. The FDI Attack procedure is described in Fig. \ref{fig:Attack_loop_inner} and each step is explained accordingly.

\subsubsection{\textbf{Collection of compromised EV measurements}}
The attack starts when enough vehicles are anticipated to connect to chargers. Once begins, the attacker uses the compromised EV measurements $S_i$ and $P_i$ to build its IEVM simulation. Since the EVs send updated measurements every $T_p$ period, attacker updates its simulation with the \emph{true} measurements of \emph{un-attacked} EVs and adds newly connected vehicles.

\subsubsection{\textbf{IEVM Simulation}}

The attacker has access to a subset of the fleet, which consists of the compromised EVs. The primary purpose of the IEVM (section \ref{IEVM}) is to emulate the dynamics of individual EVs using manipulated measurements, and it supplies these false measurements to the operator. The attacker then uses these emulated EV measurements to derive the necessary parameters for subsequent steps of the attack. As the attack progresses, the IEVMs of the targeted EVs are updated with manipulated measurements ($\tilde S_i'$, $\tilde P_i'$). In this notation, the tilde symbol ($\tilde \square $) indicates the value that is the output of the IEVM simulation, while the prime symbol ($\square '$) indicates the manipulated value. %\color{red} briefly explain what if a value having both tilde and prime?\color{olive}
The combination of both symbols ($\tilde\square '$) indicates that the measurements are sampled from the IEVM simulation, manipulated by the attacker, and then sent to the operator.

% The attacker has access to a subset of the fleet, which consists of the compromised EVs. The primary purpose of the IEVM is to simulate the expected measurement trajectory of each individual vehicle over time. %\color{red} Further explain the function of IEVM: emulate the eSSM in the operator room with manipulated measurements, so as to ensure the stealthy condition of designed attack. \color{black}
% As the attack progresses, the IEVMs of the targeted EVs are updated %\color{blue}by the attacker \color{red} who updated it? or where was it updated?\color{black}
% with manipulated
% measurements ($\tilde S_i'$, $\tilde P_i'$) \color{red} to emulate the individual EV dynamics with manipulated measurements and to emulate the eSSM model output in the operator room. The emulated ... will be further used in Step 3 to ensure the stealthiness of the designed attack vector. \color{black}
% %while their true measurements ($S_i$, $P_i$) are blocked.
% \color{blue} Here, the tilde symbol ($\tilde \square $) indicates the value that is the output of the IEVM simulation, while the prime symbol ($\square '$) indicates the manipulated value. %\color{black} For compromised vehicles that are not currently under attack, the IEVMs are continuously updated with actual SoC and power measurements to maintain accurate state estimation. When a newly compromised vehicle connects to the system, its model is added to the simulation with the most recent true measurements to ensure consistency.

%\color{blue}
At each measurement interval $T_p$, simulated IEVM measurements are employed to construct the aggregated state and the transition matrix ($\boldsymbol{\tilde{x}}(k)$ and $\tilde{A}(k)$) for the compromised vehicle subset, thereby emulating the eSSM parameters in the operator room. Furthermore, simulation is also utilized to determine each vehicle's state indices and their state transition weights, $\tilde{\boldsymbol{\pi}}_i(k)$ using the Algorithm \ref{alg:pi} to map the manipulated aggregated state distribution $\boldsymbol{\tilde{x}}'(k)$ back to individual EV measurements $\tilde S_i'$, $\tilde P_i'$.
Details will be found in \ref{FDIA Maximum Likelihood Assignment}.
% \color{olive}
% At every measurement interval $T_p$, the IEVM simulation is employed by the attacker to construct the eSSM parameters $\boldsymbol{\tilde{x}}(k)$ and $\tilde{A}(k)$ (as defined in equations \eqref{eq2} and \eqref{eq_A}, respectively \color{red} do you mean constructing $\boldsymbol{\tilde{x}}(k)$ and $\tilde{A}(k)$   by following the two equations?\color{black}).
% Furthermore, the IEVM simulation is utilized to \color{red} map the aggregated state distribution $x$ back to individual EV SOC and power value (notations)......\color{black} % determine the individual EV states,
% %$\tilde s_i(k)$ (as per equation \eqref{eq1}),
% and their corresponding state transition weights, $\tilde\pi_i(k)$, using Algorithm \ref{ag:pi}.
\color{black}
% \color{blue} added the clarification above % \color{red} The tilde symbol in the above paragraph means manipulated, does it mean the same here? \color{black}.
% A two-stage attack optimization routine (Figure \ref{fig:Attack_loop_inner}, steps 3,4,5) is then used to determine the manipulated measurements, $\tilde{S}_i'(k)$ and $\tilde{P}_i'(k)$, from the simulated measurements, $\tilde{S}_i(k)$ and $\tilde{P}_i(k)$.

\subsubsection{\textbf{Receding Horizon Attack Formulation}}
\begin{figure}[h]
    \centering
    \vspace{-10pt}
    \includegraphics[width=0.45\textwidth]{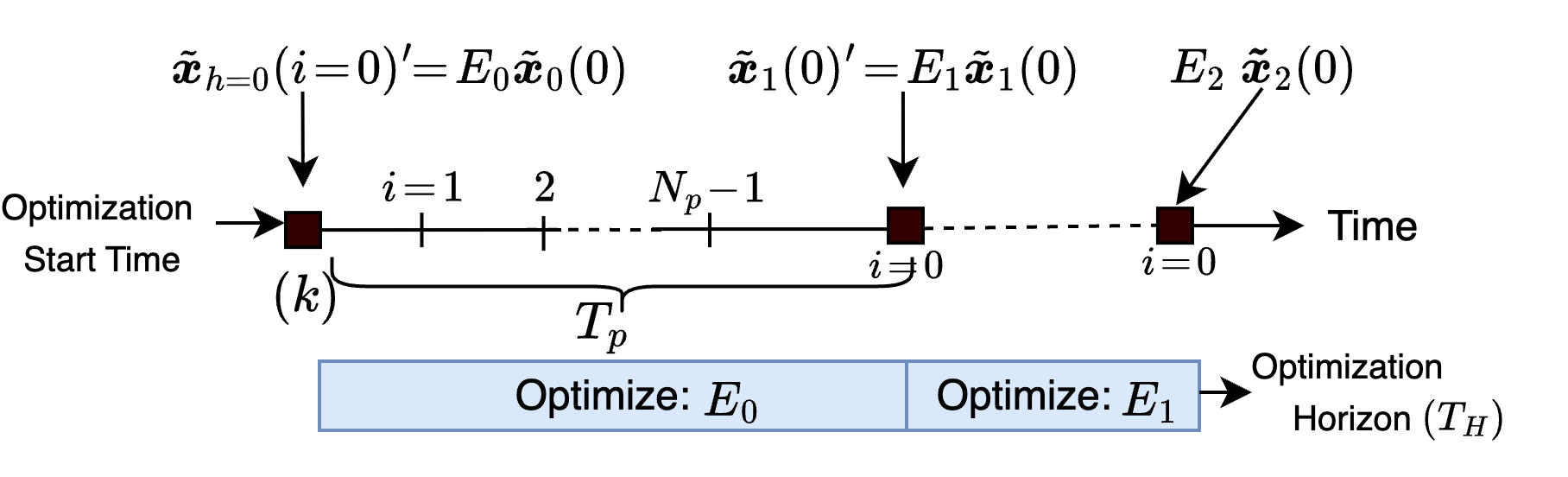}
    \caption [Illustration of the optimizations cascading timeline.] {Illustration of the optimizations cascading timeline. \par \small Output of the IEVM $\tilde{\boldsymbol{x}}_{0}(0)$ is manipulated $\tilde{\boldsymbol{x}}_{0}(0)'\!\!=\!E_0\tilde{\boldsymbol{x}}_0(0)$. At the end of period $h$, %\color{olive}
    the  manipulated state evolved under SSM %\color{red} I think it should be ``the manipulated state evolved under SSM''\color{black}
    becomes the initial state for the next period $h\!+\!1$.}
    % \vspace*{-5pt}
    \label{fig:optimization_timeline}
\end{figure}

In the first stage of the attack, $\boldsymbol{\tilde{x}}(k)$ and $\tilde{A}(k)$ are used to determine the aggregated attack vector $\boldsymbol{\tilde{x}}'$.
As shown in Fig. \ref{fig:optimization_timeline}, this stage is formulated as a multi-period optimization problem designed to find an optimal initial manipulation matrix, $E_0$ \eqref{eq:attack_1}. A receding horizon control strategy is employed where a long-term plan is formulated to inform an immediate, optimal deviation of the aggregate state vector. The attack is optimized at the current time index $k$ and planned over an optimization horizon ($T_H$).

%($k \rightarrow k+T_H.T_p$). The duration for the eSSM execution ($T_p$) multiplied by the number of times the eSSM will renew \color{red} what the difference between eSSM execution and eSSM renew. Does $T$ represent both  duration and times \color{black} ($T_H$) is the attacks optimizations horizon.
% This horizon consists of multiple $h$ eSSM renewal periods; $h \in \{0, \dots, T_H\}$, where each period, is discretized into $N_p$ time steps, indexed by $i \in \{0, \dots, N_p-1\}$. Figure \ref{fig:optimization_timeline} illustrates this timeline. %\color{red} when reading Fig. 7, not clear what $E_0\tilde{x}_0(0)$ is and others.\color{black}

% The core of the formulation is the state dynamics, where the manipulated state (at time index $k+N_p$); \color{red} what is new and old and those index relationship? In the previous section, you mention ``new'' is a corrected version of ``old'' but not manipulated.\color{blue}
% \[
% \boldsymbol{x}_{new}(k+N_p)=E_{h+1}\tilde{\boldsymbol{x}}_{h+1}(0)
% \]

% (at interval $h\!+\!1$) is determined by referencing the last estimate of the previously manipulated state
% \[
% \boldsymbol{x}_{old}(k+N_p)=E_h \tilde{\boldsymbol{x}}_h(N_p)
% \]
% Due to the stealth constraint defined in \eqref{eq:stealth_aggregate}, the state deviation over the horizon is constrained by this recursive relationship and governed by the transition matrix $A_h$:
% \begin{equation}
% \boldsymbol{\tilde{x}}_{h+1}(0) = A_h^{N_p} (E_h \boldsymbol{\tilde{x}}_h(0))
% \label{eq:attack_dynamics}
% \end{equation}

Thus, problem is formulated to find the sequence of matrices $\{E_0, \dots, E_{T_H}\}$ that maximizes the cumulative flexibility deviation over the entire planning horizon:

\vspace*{-10pt}
\begin{equation}
\max_{E_0, \dots, E_{T_H}} \quad \sum_{h=0}^{T_H} \sum_{i=0}^{N_p-1} \left[ \boldsymbol{D_l} \boldsymbol{\tilde{x}}'_h(i) - \boldsymbol{D_l} \boldsymbol{\tilde{x}}_h(i) \right]
\label{eq:attack_1}
\end{equation}
\vspace*{-5pt}

where $\boldsymbol{\tilde{x}}'_h(i) = \tilde{A}_h^i E_h \boldsymbol{\tilde{x}}_h(0)$ is the state under manipulation, % manipulated state,
$\boldsymbol{\tilde{x}}_h(i) = \tilde{A}_h^i \boldsymbol{\tilde{x}}_h(0)$ is the  state estimate without manipulation, that can be calculated from the output of IEVM.
\color{black}

The optimization is subject to:

\paragraph{Aggregate Stealth Constraint}
According to the stealth constraint \eqref{eq:stealth_aggregate}, to remain undetected, the deviation between the manipulated and reference state distributions at the start of each period must be bounded by a threshold $\epsilon$.

\vspace*{-10pt}
\begin{equation}
\| E_h \boldsymbol{\tilde{x}}_h(0) - \boldsymbol{\tilde{x}}_h(0) \| \le \epsilon \quad h \in \{0, \dots, T_H\}
\label{eq:stealth_aggregate}
\end{equation}
\vspace*{-10pt}

\noindent
where $E_h \boldsymbol{\tilde{x}}_h(0)$ represents the manipulated state distribution and $\boldsymbol{\tilde{x}}_h(0)$ the output of the IEVM, emulating part of the observation at the operator room.

\paragraph{Individual Physical Feasibility}
The manipulation must respect the physical constraints of individual EVs. The aggregate manipulation matrix $E_h$ must only represent state transitions that are feasible over a period $T_p$ (as shown in Fig. \ref{fig:StateTransitions}). Thus, the elements of the manipulation matrix is constrained as:
\vspace*{-10pt}
\begin{equation}
(E_h)_{mn} \begin{cases} \in [0, 1] & \text{if } (m,n) \in \phi \\ = 0 & \text{otherwise}  \end{cases} \quad h \in \{0, \dots, T_H\}
\label{eq:transition}
\end{equation}
\vspace*{-10pt}

where $\phi$ is the set of allowed transitions between states that satisfy individual EV power and energy constraints over the interval $T_p$. For example, if we consider state index $m=15$ (representing a SoC block of 40-49\% in the idle mode), it can only transition to idle, charging, and discharging modes within the same SoC block in $T_p$ period. %\color{olive}
Therefore, the allowed transitions for $m=15$ are $n\in \{5,15,25\}$, meaning state indexes (15,5), (15,15), and (15,25) of $\phi$ could be assigned a value between $[0,1]$. %\color{red} What does it mean by index equal to 1?\color{black}

\paragraph{Conservation Constraint}
% \color{blue}
 After state manipulation, the aggregate distribution of fleet across the state indexes must be equal to 1.
%\color{red} I think both (before and after manipulation) should be one? and this sentence doesn't seem to explain the eqn below.\color{black}\color{olive} :: Yes for both before and after; However since the sum over the aggregated state indexes before manipulation is always equates to 1, we only care about post manipulation. $E_h$ being the transition matrix (for all transitions mapping state index(s) m to n), percentage of total transitions must be add up to \%100.
% I could also write the equation as ;
\vspace*{-10pt}
\begin{equation}
\sum E_h x_h= 1
\label{eq:conservation}
\end{equation}

Although the optimization in \eqref{eq:attack_1} is solved for the entire horizon, only the initial matrix $E_0$ is implemented to generate the attack vector for the current interval:

\vspace*{-10pt}
\begin{equation}
\boldsymbol{\tilde{x}'}(k) = E_0 \boldsymbol{\tilde{x}}_0(k)
\label{eq:implementation}
\end{equation}
\vspace*{-15pt}

This is because multi-period optimization functions as a planning mechanism to determine the optimal initial manipulation ($E_0$) by forecasting its cascading effects over future periods.

\subsubsection{\textbf{Maximum Likelihood Assignment}} \label{FDIA Maximum Likelihood Assignment}

%\color{red}
% (Mimic user behavior) we are aiming to determine vehicles that have a bias towards switching to certain states given their measurement and stealthy limits.
% select whitch vehicles to do state ransitions, enure picked vehicles can atcheve that given their own phsical constraints (9-10). example 51\% EV cant go to state 60\% since ...
% to pick the vehicles to do the state transitinos so assign those higher. do the transitions sucesssfully given own physical constraints, our attack goals can be atchevd.
% we assige higher weight to ceration vehciel

%\color{blue}
Once the aggregate state distribution $\boldsymbol{\tilde{x}'}(k)$ is optimized, it must be mapped to the measurements of individual EV, as the operator receives distinct measurements from each EV rather than a single aggregate value. This second stage of the optimization performs this mapping by assigning a specific target state to each compromised vehicle using a Maximum Likelihood Estimation (MLE) approach.

Here, the attacker’s goal is to choose which vehicles’ reported values to alter so the manipulated aggregate
$\boldsymbol{\tilde x}'$ can be reconstructed by the operator while satisfying the physical constraints \eqref{eq:ind_ev_SoC_condition}, and \eqref{eq:ind_ev_P_condition}. %the intruder aim to select a set of vehicles whose reported values can be altered so that the manipulated aggregated state representation $\tilde x'$ can be recreated by the operator while meeting the physical constraints \eqref{eq:ind_ev_SoC_condition}, and \eqref{eq:ind_ev_P_condition}.
For example, an EV with $51\%$ SOC in charging mode is unlikely—given the physical power limits in \eqref{eq:ind_ev_SoC_condition} and \eqref{eq:ind_ev_P_condition}—to reach the $60–69\%$ SOC interval within $T_p$ interval, while an EV currently at $59\%$ SOC in charging mode could plausibly shift into the $60–69\%$ interval.
To capture these practical transition probabilities, a weight vector $\tilde{\boldsymbol{\pi}}_i$ (see Algorithm \ref{alg:pi} in Appendix) is computed for each EV $i$.
%In view of that, %By utilizing IEVM simulation measurements,
%a weight vector, $\tilde{\boldsymbol{\pi}}_i$ (the details can be found in Algorithm \ref{alg:pi} in Appendix), is constructed for each EV $i$, which
$\tilde{\boldsymbol{\pi}}_i$ quantifies the attacker's probabilistic assessment of the vehicle's state transitions according to their operating mode and SoC level. %This vector aims to mimic user behaviors of mode switching as well as providing misreporting of SoC measurements.
This vector is formulated as a composite of two distinct weight vectors:

% \color{red}
% In practical operational environments, EVs are susceptible to unforeseen and stochastic events, such as user interventions, intrinsic vehicle limitations, and measurement inaccuracies. These occurrences can lead to the erroneous reporting of SoC and unintended mode transitions, under the limits defined by \eqref{eq:stealth_aggregate}, \eqref{eq:ind_ev_SoC_condition}, and \eqref{eq:ind_ev_P_condition}. While SoC misreporting may arise independently of the vehicle's actual measurements, unscheduled mode switching events are more prevalent for EVs operating at either low (e.g., from Idle to Charge) or high (e.g., from Charge to Idle) SoC levels. The attacker exploit these anomalies to obscure their attack activities. Utilizing its compromised EVs IEVM measurements, a weight vector, $\tilde{\boldsymbol{\pi}}_i$ (the details can be found in Algorithm \ref{alg:pi} in Appendix), is constructed, which quantifies the attacker's probabilistic assessment of the vehicle's state transitions. This vector is formulated as a composite of two distinct weight vectors:
%the first models the vehicle's propensity for inter-mode transitions, while the second models its likelihood of transitioning between SoC intervals within a single operational mode (e.g., within Charge mode).
\color{black}

% In real world conditions, unexpected and random events (e.g., user intervention, vehicle internal limits, measurement errors) of EVs misreporting SoC measurements and switching modes without the directive of the operator is expected under the conditions of \eqref{eq:stealth_aggregate}, \eqref{eq:ind_ev_SoC_condition} and \eqref{eq:ind_ev_P_condition}. While misreporting of SoC may occur independent of EV's true measurements, unexpected mode switching events are more likely to occur for EVs in low SoC levels (e.g., from Idle to Charge) and high SoC levels (e.g., from Charge to Idle). The attacker leverages such anomalies to hide its attack. Given IEVM measurements of compromised EVs, it constructs a weight vector $\tilde{\boldsymbol{\pi}}_i$ (the details can be found in Algorithm \ref{ag:pi} in Appendix) which reflects the attacker's probabilistic assessment of the vehicle's state transitions. Thus, this weight vecor is  modeled as the combination of two distinct weight vectors: the first weight vector models a vehicle's likelihood of transitioning between different operating modes, while the second models its likelihood of transitioning between SoC intervals within the same operating mode (e.g., in Charge).
% % \color{red} what's the rationale\color{black}

\paragraph{State Transition Across Operating Modes} These model the likelihood of vehicle changing operating modes.  EVs with higher SoC while in Charging Mode (CM) are assigned larger weights for transitioning to Idle Mode (IM), which facilitates  $\text{CM} \rightarrow \text{IM}$. Conversely, vehicles with lower SoC are assigned higher weights for following the $\text{DM} \rightarrow \text{IM} \rightarrow \text{CM}$ sequence. %Models vehicles chance to transition its mode by assigning higher weights for vehicles with higher SoC values in charging mode, to switch to idle mode, thus facilitating the transition $\text{CM} \rightarrow \text{IM}$. Conversely, it assigns higher weights for vehicles with lower SoC values to transition through the sequence $\text{DM} \rightarrow \text{IM} \rightarrow \text{CM}$.

\paragraph{State Transition Within an Operating Mode} These capture the probability of advancing within the same operating mode to the next SoC interval. An EV in CM with a high local SoC position %within its discrete interval
receives greater weight for moving to the next SoC interval. Conversely, IM or DM with low SoC position receives greater weight for moving to the previous SoC interval.
\color{black} %enabling a gradual, and feasible shift in undetectable shift in perceived distribution.  %Models vehicles chance to transition its state with in the same mode while adhere to defined stealthy conditions. %\color{black}This is based on the premise that, in the absence of an explicit operator control signal or entry into a special state, the EV will sustain its current operating mode.
%an EV in CM with higher $SoC_{loc,i}$ is weighted more for advancing to the next SoC interval, allowing a gradual, undetectable shift in perceived distribution. This is facilitated by intra-state state-of-charge ($SoC_{loc}= mod(S,N_s)$), which quantifies an EV's precise SoC position within its discrete interval (e.g., an EV at 58\% SoC in a 50-59\% interval has $SoC_{loc,i}=0.8$. Assume SoC granularity $G_{SoC}=1\%$. %This enables subtle manipulation: an EV in CM with higher $SoC_{loc,i}$ is weighted more for advancing to the next SoC interval 60-69\%, allowing a gradual, undetectable shift in perceived distribution.

\color{black}

% \color{red} How is $\pi$ determined? You may need to briefly explain how $\pi$ is determined for different cases. \color{black}

The attacker's objective is to determine the optimal state transition for each compromised EV, leveraging $\tilde{\boldsymbol{\pi}}_{i...N_{comp}}$ to guide strategic switching decisions. Thus, formulating a mixed integer linear programming problem where the decision variables $\theta_{i, j} \; \forall j \!\in\! 3N_s$ (excluding special states) represent the state index transitions for each EV.

\vspace*{-8pt}
\begin{equation}
\label{eq:attack_2}
\max_{\boldsymbol{\theta}}  \quad \sum_{i=1}^{\mathcal{N}_{comp}} \sum_{j=1}^{3N_s} \theta_{ij} \log(\tilde{\boldsymbol{\pi}}_i(j \mid S,P))
\end{equation}

\noindent
subject to:
\vspace*{-10pt}
\[
    \theta_{ij} = 0 \quad  \forall i \in \mathcal{N}_{comp} \quad \forall \tilde{\boldsymbol{\pi}}_i(j \mid S,P) \;\leq\; 0
\]
\vspace*{-10pt}
\[
    \sum_{j=1}^{3N_s} \theta_{ij} = 1 \quad  \forall i \in \mathcal{N}_{comp}(k)
\]
\vspace*{-10pt}
\[
    \sum_{i=1}^{\mathcal{N}_{comp}} \theta_{ij} = \mathcal{N}_{comp}(k) \boldsymbol{\tilde{x}}'_j \quad \forall j \in 3N_s
\]
\vspace*{-10pt}

\noindent
where $\mathcal{N}_{comp}(k)$ is the number of compromised EVs and  $\tilde{\boldsymbol\pi}_i(j \mid S,P)$ is the assigned weight of EV $i$ transition to state at index $j$, given its measurements.
% \color{red}I think it is a mixed integer linear program that can be solved analytically? \color{black}

\subsubsection{\textbf{State Index to Measurement Mapping}}

Once the optimal state assignment $\boldsymbol{\theta}_i$ is determined for each compromised vehicle $i$, the final step maps this abstract assignment to concrete manipulated measurements ($\tilde{S}_i', \tilde{P}_i'$). The one-hot vector $\boldsymbol{\theta}_i$, where $\theta_{ij}=1$, identifies the target state index $j$ for the next time step, defining the new operating mode (e.g., Charging, Discharging, or Idle) and the corresponding SoC range.

\vspace*{-8pt}
\begin{equation}
\label{eq:state_to_P_map}
\small
\tilde{P}_i' :
  \begin{cases}
    0  &\text{if } \text{CM} \rightarrow \text{IM} \\
    0  &\text{if } \text{DM} \rightarrow \text{IM} \\
  \end{cases}
\quad
\tilde{P}_i' :
  \begin{cases}
    P_{d,i}  & \text{if } \text{IM} \rightarrow \text{DM} \\
    P_{c,i}  & \text{if } \text{IM} \rightarrow \text{CM}
  \end{cases}
\end{equation}

\vspace*{-5pt}
\begin{equation}
\label{eq:state_to_SoC_map}
\tilde{S}_i' = \tilde{S}_i + sgn(\tilde{P}_i') \cdot \Biggl \lceil \frac{ |\tilde{P}_i' \cdot T_p |}{Q_i}  \Biggl \rceil
\end{equation}

This process ensures that the generated measurements ($\tilde{S}_i', \tilde{P}_i'$) are consistent with the vehicle's physical capabilities and adhere to the stealthy conditions in \eqref{eq:ind_ev_SoC_condition} and \eqref{eq:ind_ev_P_condition}.

\subsubsection{\textbf{Update IEVM with Manipulated Measurements}} All manipulated measurements are used to update the IEVM, which will be used in the next attack optimization interval.

\subsubsection{\textbf{IEVM's Transmission of Manipulated Measurements}} Updated simulation measurements of compromised EVs are sent to the operator.

\subsubsection{\textbf{eSSM Construction}} To construct its eSSM model, the operator utilizes the most recent EV measurements, which consist of both true ($S_i$, $P_i$) and manipulated ($\tilde S_i'$, $\tilde P_i'$) values. %\color{red} \st{I am not sure about this. need discussion.}\color{blue} \st{Assume $\alpha$ being the fraction compromised vehicles, operators aggregated state representation can be written as the linear combination of safe $\bar x$ and manipulated $\tilde{x}'$ aggregated state vectors;}

%\begin{equation}
%    \label{eq:operators_state_rep}
%    \hat{x} \approx \alpha \tilde{x}' + (1-\alpha)\bar x
%\end{equation}
%\color{black}
\section{Numerical Validation}
This section details the numerical validation of the proposed stealthy FDIA, quantifying its impact on V2G system operations, including grid flexibility estimation and frequency stability.
% Simulations were conducted using a custom environment modeling EV fleet \color{red}(Python) \color{black} and power grid interactions \color{red}(MATLAB)\color{black}
Simulations were conducted with Python and MATLAB.

The simulation setup involves a fleet size of 10,000 EVs, with their comprehensive specifications detailed in Tables \ref{tab:traveling_parameters} \cite{8764460}. The utility's eSSM utilizes 10\% SoC interval discretization. To simulate a realistic attack scenario, a random 30\% of the connected EVs are designated as compromised at the simulation's onset. These compromised EVs are then subject to the attacker's manipulation of their reported measurements immediately after sending their initial connection signal to the operator.

\begin{table}
    \centering
    \vspace*{-10pt}
    \caption{Fleet Parameters}
    \scriptsize
    \label{tab:traveling_parameters}
    \begin{tabular}{c c c}
        \hline
        \textbf{Parameter} & \textbf{Description} & \textbf{Value*} \\
        \hline
        $S_{s}$ & Start Charging SOC & $N(0.2, 0.05) \in [0.2, 0.4]$ \\
        $S_{d}$ & Demanded SOC for Travel & $N(0.85, 0.03) \in [0.75, 0.95]$ \\
        $t_{s}$ & Start Charging Time (h) & $N(-6.5, 3.4) \in [0, 5.5]$ \\
        $t_{f}$ & Finish Charging Time (h) & $N(8.9, 3.4) \in [0, 20.9]$ \\
        $S_{min}/S_{max}$ & Minimum/Maximum SOC Value & $0.05/0.95$ \\
        $P_c/P_d$ & Charging/Discharging Power (kW) & U(6.0, 8.0) \\
        $\eta$ & Charging/Discharging Efficiency & U(0.88, 0.95) \\
        $Q$ & Battery Capacity (kWh) & U(20.0, 40.0) \\
        $T_p$ & Optimization Period & 5 min.\\
        $T$ & Optimization interval count & 20 sec.\\
        $\epsilon$ & Aggregated State Distribution Error & 0.01\\
        $T_H$ & Optimization horizon & 2\\
        $N_p$ & Optimization interval count & 5\\
        \hline
    \end{tabular}
    \begin{tablenotes}
        \item[*] $U(a, b)$ denotes a uniform distribution with variation range $[a, b]$.
        \item[*] $N(\mu, \delta)$ denotes a normal distribution where $\mu$ is the mean value and $\delta$ is the standard deviation. $[a, b]$ is the variation range.
    \end{tablenotes}
    \vspace*{-20pt}
\end{table}

\subsection{FDIA with No Control}
\begin{figure}[H]
    \centering
    \vspace*{-13pt}
    \includegraphics[width=0.50\textwidth]{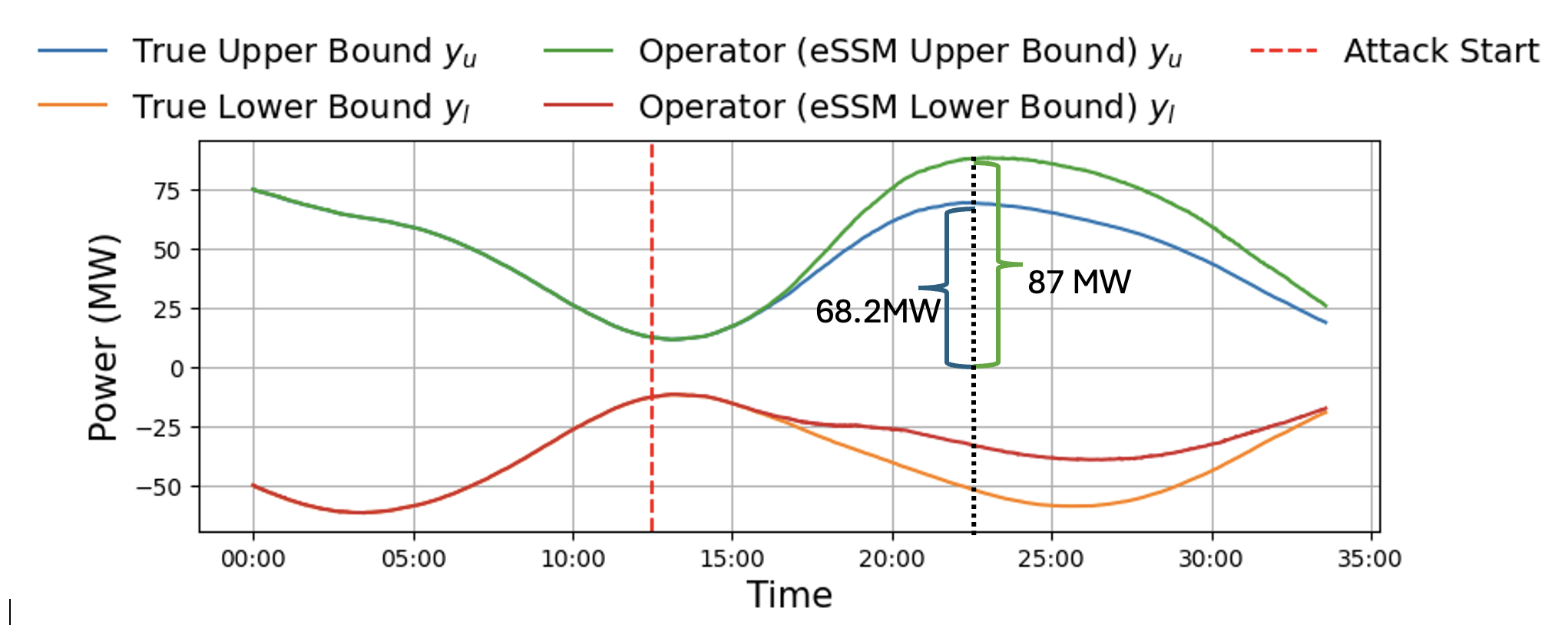}
    \caption[Grid Flexibility with No Control]{Grid Flexibility with No Control. \quad \small Upper and lower bounds of EVs' flexibility from SSM under attack % grid flexibility
    are represented as $y_{u}$ and $y_{l}$, respectively.}
    \label{fig:ssm_w_no_feedback}
\end{figure}

In this scenario, the operator solely utilizes the eSSM to estimate the grid's inherent flexibility, without actively dispatching control signals. As demonstrated in Fig. \ref{fig:ssm_w_no_feedback}, upon attack initiation at 12:00, the eSSM's estimated grid flexibility gradually diverges from its true value. The marked overestimation of the upper bound of the eSSM's flexibility is a direct consequence of the compromised EVs being manipulated to continually report their mode as charging, thereby inflating the perceived available power for grid services. This section highlights the attacker's ability to create a persistent illusion of greater grid support.

\vspace*{-20pt}
\subsection{FDIA with Control}

This subsection evaluates the attack's efficacy when the utility actively employs the eSSM for frequency regulation within secondary frequency control under a closed-loop framework. As depicted in Fig. \ref{fig:Communication Flow} and described in Section \ref{Preliminaries}, the operator generates control signals based on its eSSM to meet dispatch requests.

\begin{figure}[H]
    \centering
    \vspace*{-10pt}
    \includegraphics[width=0.50\textwidth]{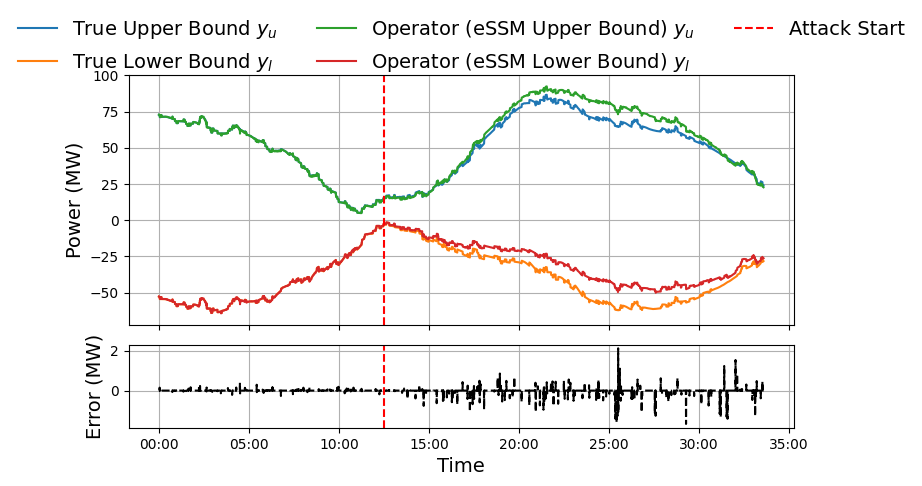}
    \caption[Grid Flexibility with Control]{Grid Flexibility with Control %\color{red} modifications for legends are needed as well\color{black}
    }
    \label{fig:ssm_w_feedback}
    \vspace*{-12pt}
\end{figure}

The attacker manipulates the compromised EV measurements, causing the operator to perceive the aggregated state distribution as \eqref{eq:implementation}. %\color{red} \st{Do you still want that eqn? } \color{black}
Consequently, the operator broadcasts faulty control signals intended to adjust the EVs' state indices. However, since the EVs' true measurements ($P_i,S_i$) are unaltered, they disregard these control signals, which do not correspond to their actual state indices. As illustrated in Fig. \ref{fig:ssm_w_feedback}, this discrepancy results in a substantial degradation in power correction accuracy. %\color{red} \st{I think the EVs don't report inaccurate states. Intruder does. Check Fig. 5, the description here may need adjustment .Since no control were discussed in previous section. It may need some description about how this works. For example in Fig. 5, how the control works.} \color{black}

Quantitatively, the attack leads to a dramatic increase in the Mean Absolute Percent Error (MAPE) of power correction. The pre-attack MAPE of 5.49\% surges to 520.28\% during the attack period. This significant increase demonstrates the attack's success in undermining the operator's capacity to accurately manage grid services. This would directly translate into severe operational inefficiencies, substantial economic losses, and critically compromise the reliability and stability of the grid.

% \begin{table}[h!]
%     \centering
%     \caption{Optimization Parameters}
%     \scriptsize
%     \label{tab:attack_parameters}

%     \begin{tabular}{c c c}
%         \hline
%         \textbf{Parameter} & \textbf{Description} & \textbf{Value*} \\
%         \hline
%         $\epsilon$ & Aggregated State Distribution Error & 0.01\\
%         $T_H$ & Optimization horizon & 2\\
%         $N_p$ & Optimization interval count & 5\\
%         \hline
%     \end{tabular}
% \end{table}

\subsection{AGC Simulation}

\begin{figure}[h]
    \centering
    \vspace*{-5pt}
    \includegraphics[width=0.48\textwidth]{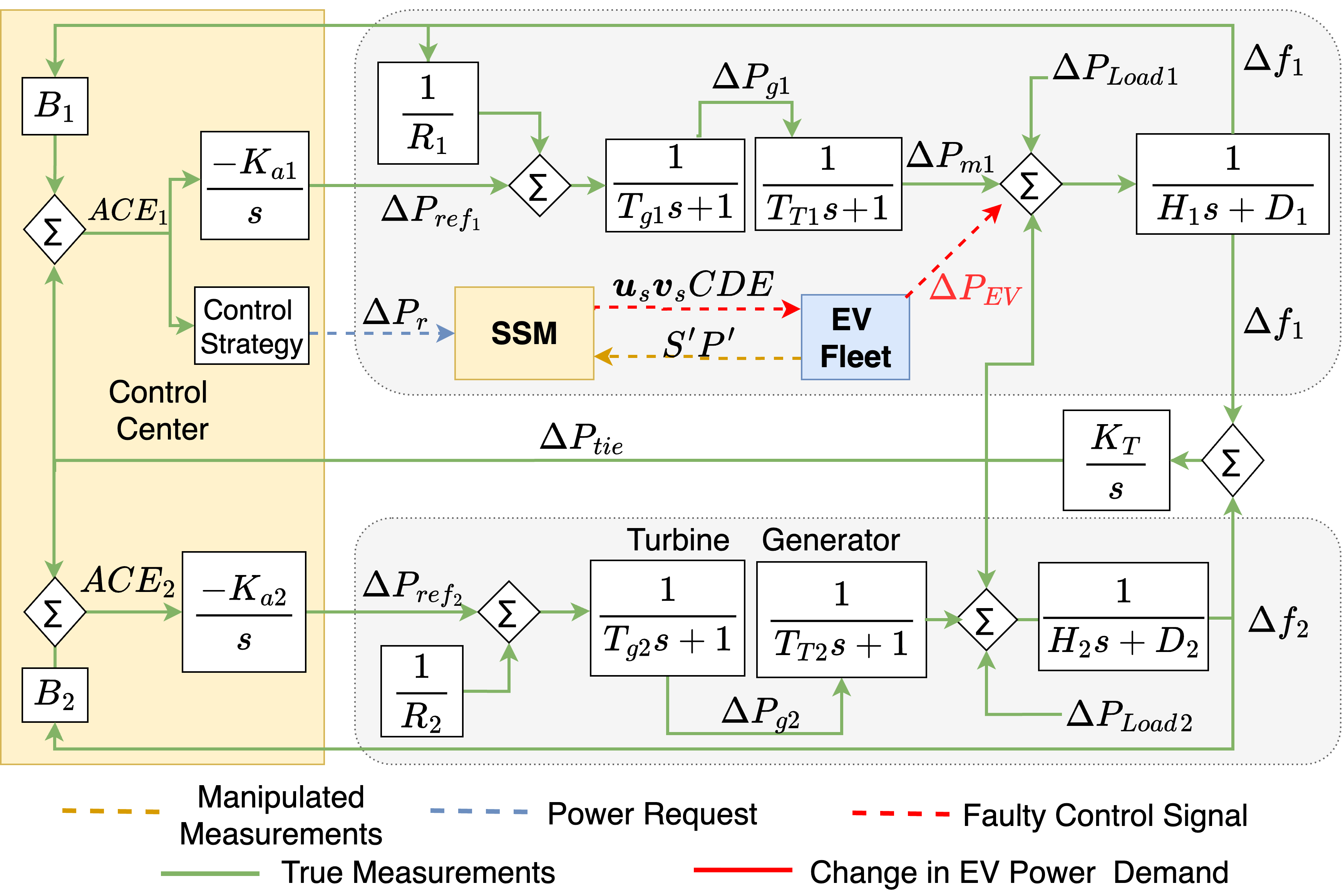}
    \caption{2-Area AGC Simulation With Aggregated EV Model}
    \label{fig:AGC}
    \vspace{-10pt}
\end{figure}

To illustrate the critical real-world implications of the misestimated flexibility, a 2-area AGC system was simulated using Simulink. The AGC system as shown in Fig. \ref{fig:AGC} and parameters are detailed in Table \ref{tab:agc_parameters}. Both generators are configured to have a maximum operating point of 200 MW, while the base load on both areas is $P_{L}=0.475$ p.u. The operator in Area 1 possesses the V2G capability and can perform power corrections by dispatching control signals. %\color{olive} Removed sentence  \color{red} what does this sentence mean? \color{black}

A predefined control strategy is implemented, designed to receive the  Area
Control Error (ACE) and determine the requisite power change ($\Delta P_r$) for frequency correction. We simulated a case where the power demand in Area 1 sharply increases by %\color{blue}
50 MW at 22:00 (bounds of the grid  flexibility as shown in Fig. \ref{fig:ssm_w_no_feedback}), %\color{olive} I forgot to update this section after modifying the figures \color{red} I don't see there is 100MW at 21:00 from Fig. 7. \color{black}
represented by a step function $\Delta P_{L1}$. Given that the conventional generators are unable to compensate for this demand entirely, the control center dispatches $\Delta Pr=50$ MW to the EV operator. %\st{\color{red} Again, I don't see there is enough 85MW from Fig. 8. Should I look Fig. 7 or Fig. 8? Also, can you mark the value of flexibility at 22:00 in the figure? \color{black}}

The true upper bound of grid flexibility (Fig. \ref{fig:ssm_w_no_feedback}), denoted as $y_u$ (measured at $68.2\text{ MW}$), is technically sufficient to satisfy the utility's power request ($50\text{ MW}$). However, the EV operator generates the control signals based on a manipulated aggregated state, $\boldsymbol{\tilde{x}'}$. The construction of these control signals relies on a priority function (detailed in Section \ref{Preliminaries}), which is fundamentally driven by the aggregated state distribution. Consequently, any significant deviation in this distribution leads to the generation of faulty control signals. In the event described, the EVs intended for state-switching are not located within the presumed aggregated states. As a result, the targeted EVs discard the received signals because they are deemed non-applicable, leading to a substantial reduction in performance. Therefore, the actual power contribution delivered by the fleet, $\Delta P_{EV}$, is only $25.95\text{ MW}$.
%\color{blue}

%\color{red} \st{I don't understand the reason given here. I think it is because the user doesn't follow the control given their physical constraints, so only 63.7 MW was contributed. It should not because that the control signals are designed based on the attacker's manipulated measurements. Please check your descriptions.}\color{blue}

\begin{figure}
     \centering
     \vspace{-13pt}
     \includegraphics[width=0.45\textwidth]{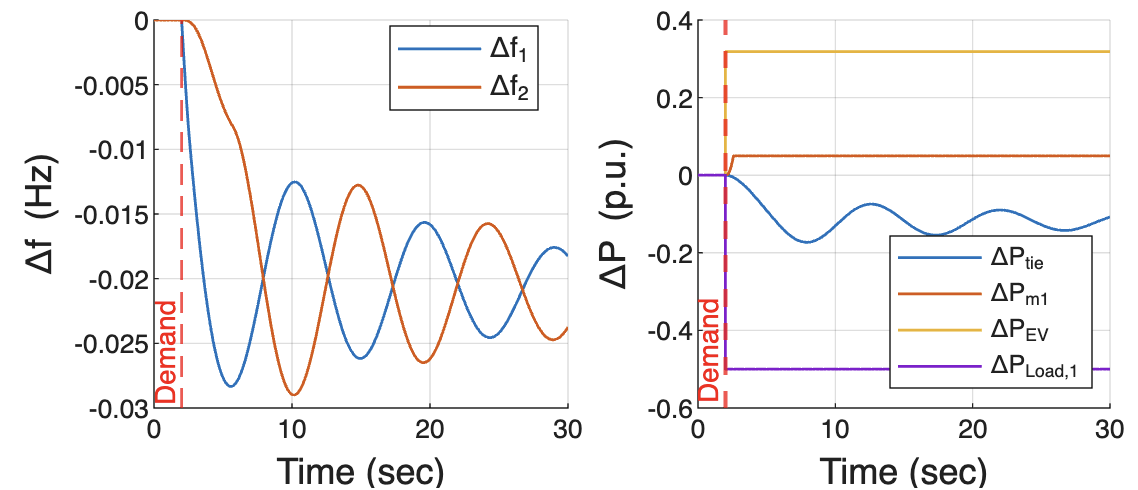}
     \label{fig:AGC_results}
     \vspace{5pt}
    \caption{Impact of FDIA on $\Delta f$ and $\Delta P$ at 22:00 based on the flexibility shown in Fig. \ref{fig:ssm_w_no_feedback}. \quad \small $\Delta P_{EV}$ is dispatched 100 ms after the load increase. Negative $\Delta P_{tie}$ indicates the direction of the power flow (Area-2 to Area-1).}
    \label{fig:AGC_results}
    \vspace{-5pt}
\end{figure}

The mismatch between perceived and actual supply leads to a critical demand-supply imbalance of 4.05MW. As a direct consequence, the power flow over the tie line increases to provide emergency support to the affected area, and subsequently, significant frequency deviations are observed in both areas, as represented in Fig. \ref{fig:AGC_results}. %This result underscore the direct threat that such stealthy cyberattacks pose to grid frequency stability and overall power system reliability, even when direct physical control is not achieved.

\begin{table}
    \centering
    \vspace{-10pt}
    \caption{AGC Parameters}
    \scriptsize
    \label{tab:agc_parameters}
\scriptsize
\begin{tabular}{|c|c|c|c|c|c|c|c|}
    \hline
    Parameters & $H_i$ & $D_i$ & $R_i$ & $T_{gi}$ & $T_{Ti}$ & $K_{ai}$ & $B_i$ \\
    \hline
    Area 1 & 10 & 0.6 & 0.05 & 0.2 & 0.5 & 0.3 & 20.6 \\
    \hline
    Area 2 & 8 & 0.9 & 0.0625 & 0.3 & 0.6 & 0.3 & 16.9 \\
    \hline
    \multicolumn{5}{|l|}{$K_{T} = 2$} & \multicolumn{3}{l|}{Base Power = 400 MVA} \\
    \hline
\end{tabular}
\vspace*{-15pt}
\end{table}
\vspace*{-10pt}

\section{Conclusion}

This paper proposed a novel stealthy FDIA
model targeting V2G coordination based on eSSM. Under the realistic assumption of a limited adversary capable of compromising only a subset of EVs, the proposed attack demonstrates how manipulated State of Charge (SoC) and power measurements can deceive the operator’s flexibility estimation while remaining consistent with eSSM expectations. Simulation results revealed that the proposed FDIA can significantly distort the perceived flexibility, leading to incorrect control actions and degraded frequency stability, even without direct access to the control infrastructure. These findings reveal that even a partial, stealthy compromise of EV data can lead to severe operational impacts, underscoring the urgent need for robust anomaly detection and defense strategies in future V2G deployments.
%The key takeaway is that even a limited compromise of EV measurements, when executed stealthily, can lead to severe operational consequences, highlighting an urgent need for robust anomaly detection and defense mechanisms in future V2G deployments. %These findings highlight the potential risks of relying solely on aggregated model-based coordination in V2G systems.

Future work will focus on developing dynamic, feedback-aware attack and detection models, along with resilient state estimation mechanisms capable of identifying stealthy, coordinated data manipulation within large-scale, aggregated V2G frameworks.

\vspace{-10pt}
\section{Appendix}
\vspace{-10pt}

\begin{algorithm}
\small
    \caption{Transition Weights $\boldsymbol{\pi}(S, P)$}
    \label{alg:pi}
    % Use \tcp for comments (stands for 'text comment pointer')
    \SetKwComment{Comment}{// }{}

    % Combined initializations
    $\boldsymbol{\beta}_j \leftarrow 0.8(1 - j^2/100)$; $\boldsymbol{\beta}_j' \leftarrow 0.8 - \boldsymbol{\beta}_j$, $\forall j \in \{1,..,N_s\}$\;

    \For{$i=0$ \KwTo $\mathcal{N}_{comp}$}{
        $w_{stay} \leftarrow 0.8$; $w_{D(own)}, w_{U(p)}, w_{R(ight)}, w_{L(eft)} \leftarrow 0$\;
        $SoC_{loc} \leftarrow \text{mod}(S, N_s)$; $j^* \leftarrow f(S, P)$\;

        \If{$j^* > 3N_s$}{ \Comment{Continue}}

        \uIf{$P > 0$}{
            $w_{U} \leftarrow \beta'_{j^*}$\;
            \If{$SoC_{loc} \ge 0.9$}{$w_{R} \leftarrow 0.9$\;}
        }
        \ElseIf{$P = 0$}{
            $w_{D} \leftarrow \beta_{j^*}$\;
            \If{$SoC_{loc} \ge 0.9$}{$w_{R} \leftarrow 0.2$\;}
            \ElseIf{$SoC_{loc} \le 0.1$}{$w_{L} \leftarrow 0.2$\;}
        }
        \ElseIf{$P < 0$}{
             $w_{D} \leftarrow \beta_{j^*}$\;
            \If{$SoC_{loc} \le 0.1$}{$w_{L} \leftarrow 0.9$\;}
        }
        $\mathbf{w}_{raw} \leftarrow [w_{stay}, w_{D}, w_{U}, w_{R}, w_{L}]$\;
        $\boldsymbol{\pi}_i \leftarrow \sigma(\mathbf{w}_{raw})_i$\;
    }
\vspace*{-5pt}
\end{algorithm}
\vspace*{-5pt}

%\color{red}
\noindent Algorithm \ref{alg:pi} is employed to assign state index transition weights to each compromised and connected EV $\mathcal{N}_{comp}$ according to their SoC and power measurements.
As depicted in  Fig. \ref{fig:StateTransitions}, EVs state index transition are limited to their neighboring indexes thus, all possible transition stay, right, left, up, down are assigned with a weight ($w_{\text{stay}}$, $w_{\text{R}}$, $w_{\text{L}}$, $w_{\text{U}}$ ,$w_{\text{D}}$), while invalid state transition weights are set to zero. The resulting raw weights, denoted $\mathbf{w}_{\text{raw}} \in \mathbb{R}^{3N_{s}}$,
are values ties to each available state index transition. These values are then normalized normalized with \emph{softmax} function $\sigma(\mathbf{w})_i = \frac{e^{w_i}}{\sum_{j=1}^{N_{comp}} e^{w_j}}$ for each EV $i$ to convert arbitrary transition weights to probability distribution. Assume $j^*$ is the current state index of the EV.

Except the special states $j^*>3N_s$ (explained in \ref{The Extended State Space Model of Aggregated EVs}), all states indexes are assigned multiple transition weights, allowing the attacker to manipulate EVs state index by changing their measurements.  As explained in section \ref{FDIA Maximum Likelihood Assignment}, while weights for transitions across operating modes are determined according to EV's SoC interval ($\beta_j$ and $\beta_j'$), transitions within an operating mode are determined by EV's local SoC position ($SoC_{loc}$).
\color{black}

% \noindent Algorithm \ref{alg:pi} is employed to assign weights for each valid state transition stay, right, left, up, down respectively ($w_{\text{stay}}$, $w_{\text{R}}$, $w_{\text{L}}$, $w_{\text{U}}$ ,$w_{\text{D}}$) while invalid state transition weights are set to zero as indicated in Fig. \ref{fig:StateTransitions}. The resulting raw weights, denoted $\mathbf{w}_{\text{raw}} \in \mathbb{R}^{3N_{s}}$, are then normalized with \emph{softmax} function $\sigma(\mathbf{w})_i = \frac{e^{w_i}}{\sum_{j=1}^{N_{comp}} e^{w_j}}$ for each EV $i$ to convert arbitrary transition weights to probability distribution. Assume $j^*$ is the current state index of the EV.

\begin{itemize}
\small
\item $\boldsymbol{\beta}$: $\text{DM} \!\rightarrow \! \text{IM}\! \rightarrow \!\text{CM}$:  EVs with lower SoC are prioritized to falsely indicate mode switching.
\item $\boldsymbol{\beta}'$: $\text{CM} \!\rightarrow\! \text{IM}$: EVs with higher SoC values are assigned higher weights for switching to IM.
\end{itemize}
\vspace*{-10pt}

\bibliographystyle{IEEEtran}
\bibliography{reference.bib}
\end{document}